\documentclass[aps,prb,twocolumn,showpacs,superscriptaddress]{revtex4}
\usepackage{epsfig}
\usepackage{graphicx}
\usepackage{amsfonts}
\usepackage[figuresright]{rotating}
\usepackage{amssymb}
\usepackage{amsmath}
\usepackage{dcolumn}
\usepackage{bm}

\def\be{\begin{equation}} \def\ee{\end{equation}}
\def\bea{\begin{eqnarray}} \def\eea{\end{eqnarray}}

\def\nn{\nonumber}
\def\pp{\parallel}

\begin{document}
\title{Frustrated Bose-Einstein condensates with non-collinear orbital 
ordering}
\author{Zi Cai}
\email{zcai@physics.ucsd.edu}
\affiliation{Department of Physics, University of California, San
Diego, CA92093}
\author{Yu Wang}
\email{yu.wang@whu.edu.cn}
\affiliation{School of Physics and Technology, Wuhan University,
Wuhan 430072, China}
\author{Congjun Wu}
\affiliation{Department of Physics, University of California, San
Diego, CA92093}
\affiliation{School of Physics and Technology, Wuhan University,
Wuhan 430072, China}

\begin{abstract}
We investigate the unconventional Bose-Einstein condensations with the 
orbital degree of freedom in the 3D cubic optical lattice, which gives rise 
to various exotic features absent in conventional scalar and spinor 
Bose-Einstein condensations.
Orbital angular momentum moments are formed on lattice sites breaking
time-reversal symmetry spontaneously.
Furthermore, they exhibit orbital frustrations and develop a chiral 
ordering selected by the ``order-from-disorder'' mechanism.
\end{abstract}
\maketitle

Bose-Einstein condensation (BEC) is a striking phenomenon of quantum
many-body systems, characterized by a uniform phase that spontaneously
breaks the $U(1)$ symmetry.
By introducing extra degrees of freedom, novel quantum condensates
with even more exotic symmetry breaking patterns and topological
structures emerge.
A familiar example is the superfluid $^3$He phases, which is a spin-triplet
$p$-wave Cooper pairing condensate with both spin and orbital degrees of
freedom \cite{Leggett1975,Vollhardt1990}.
It exhibits a variety of rich structures that simultaneously incorporate
the symmetries of liquid crystals, magnets and scalar superfluids.
Consequently, the superfluid $^3$He systems possess fundamental connections
with particle physics, and exemplifies fundamental concepts of
modern theoretical physics \cite{Volovik2003}.

The rapid developments of cold atom gases provide an opportunity
other than $^3$He to explore exotic condensations with internal degrees of
freedom. Spinor atomic gases, composed of atoms with hyperfine spins,
simultaneously exhibit magnetism and superfluidity
\cite{Kurn1998,Stenger1998,Ho1998}. 
Furthermore, orbital is a degree of freedom independent of spin and charge.
It is originally investigated
in condensed matter transition metal oxides, which plays an
important role in superconductivity, metal-insulator transition, and
quantum magnetism \cite{Imada1998, Tokura2000}. 
Introducing orbital into cold atom gases has been theoretically 
investigated, which leads to unconventional BECs with complex-valued condensed
wavefunction of bosons and spontaneously breaking of time-reversal
symmetry \cite{Liu2006,Isacsson2005,Kuklov2006,Wu2006,Wu2009a,Cai2011,
Li2012,Collin2010}.
Excitingly, the recent experimental progress has realized the
meta-stable BECs in high-orbital bands exhibiting complex-valued
condensate wavefunctions at non-zero wavevectors and orbital
orderings \cite{Mueller2007,Wirth2011,Wirth2011a}.

In this paper, we investigate the properties of an orbital BEC in the
$p$-orbital bands of a cubic optical lattice.
Although orbital BECs share many properties with the ferromagnetic
phase in spinor BECs, there are crucial differences between them.
In spinor BECs, the internal degrees of freedom of hyperfine spin degree
is not coupled to the lattice.
However for orbital BECs, the ordering of the orbital angular momentum
comes from the atom motion within each optical site,
and thus is closely related to the motion of atoms in optical lattice.
As we will show, the uniqueness of the orbital degree of freedom gives
rise to a whole host of exotic phenomena, such as orbital frustration
and concomitant non-collinear orbital orderings selected by the
{\it ``order-from-disorder''} mechanism.
The selected ordering pattern exhibits an orbital angular momentum
moment chirality.

The Hamiltonian of bosons pumped to the $p$-orbital bands of a cubic
optical lattice is described by a multi-orbital Bose-Hubbard model,
$H=H_t+H_{int}$, \bea
H_t&=&\sum_{\vec{r}\mu\nu}[t_\parallel\delta_{\mu\nu}
-t_\perp(1-\delta_{\mu\nu})]
\Big\{ p^\dag_{\mu,\vec{r}+a\vec{e}_\nu}
p_{\mu\vec{r}}+h.c. \Big\}, 
\nn \\
H_{int} &=& \frac U2 \Big (n_{\vec{r}^2}-\frac{1}{3}
\vec{L}_{\vec{r}}^2 \Big ), \label{Eq:kin} \eea where
$\mu,\nu=x,y,z$ denote the orbital indices, $a$ is the lattice
constant. $p_{\mu,\vec{r}}$ ( $p_{\mu,\vec{r}}^\dag$) are
annihilation (creation) operators for bosons at  site $\vec{r}$ in
orbital $\mu$. 
$n_{\vec{r}}$ is the total particle number operator and $\vec{L}_{\vec{r}}$ 
represents the total orbital angular momentum on site $\vec{r}$. 
$t_\parallel$ and $t_\perp$ describe the
nearest-neighbor hopping matrix elements along the longitudinal and
transverse directions, respectively. Using the terminology of
chemistry, they are denoted as $\sigma$ and $\pi$-bonding,
respectively. Due to the odd parity of $p$-orbitals, $t_\parallel$
and $t_\perp$  are positive. The strong anisotropy of the $p$-band
Wannier wave-function implies that $t_\perp\ll t_\parallel$. The
onsite interaction term, $H_{int}$, reflects the Hund's type physics
generalized to bosons, {\it i.e.}, bosons prefer to occupy
complex-valued orbitals of the $p_{\hat e_1}+ip_{\hat e_2}$ type
with $\hat e_1 \perp \hat e_2$. This complex-valued orbital has
larger spatial extension than the real ones of $p_\mu$, and thus the
repulsive interactions are reduced and simultaneously the onsite
orbital angular momenta are maximized \cite{Liu2006,Wu2009a}.

\begin{figure}[htb]
\includegraphics[width=0.35\linewidth]{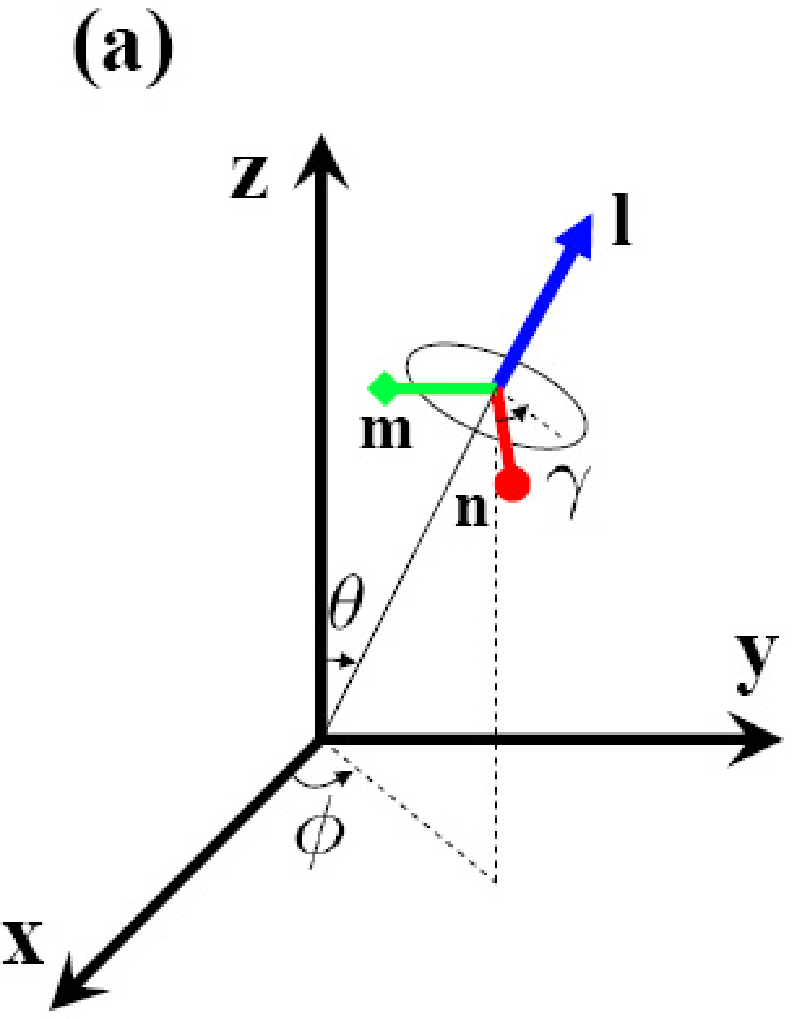}
\includegraphics[width=0.44\linewidth]{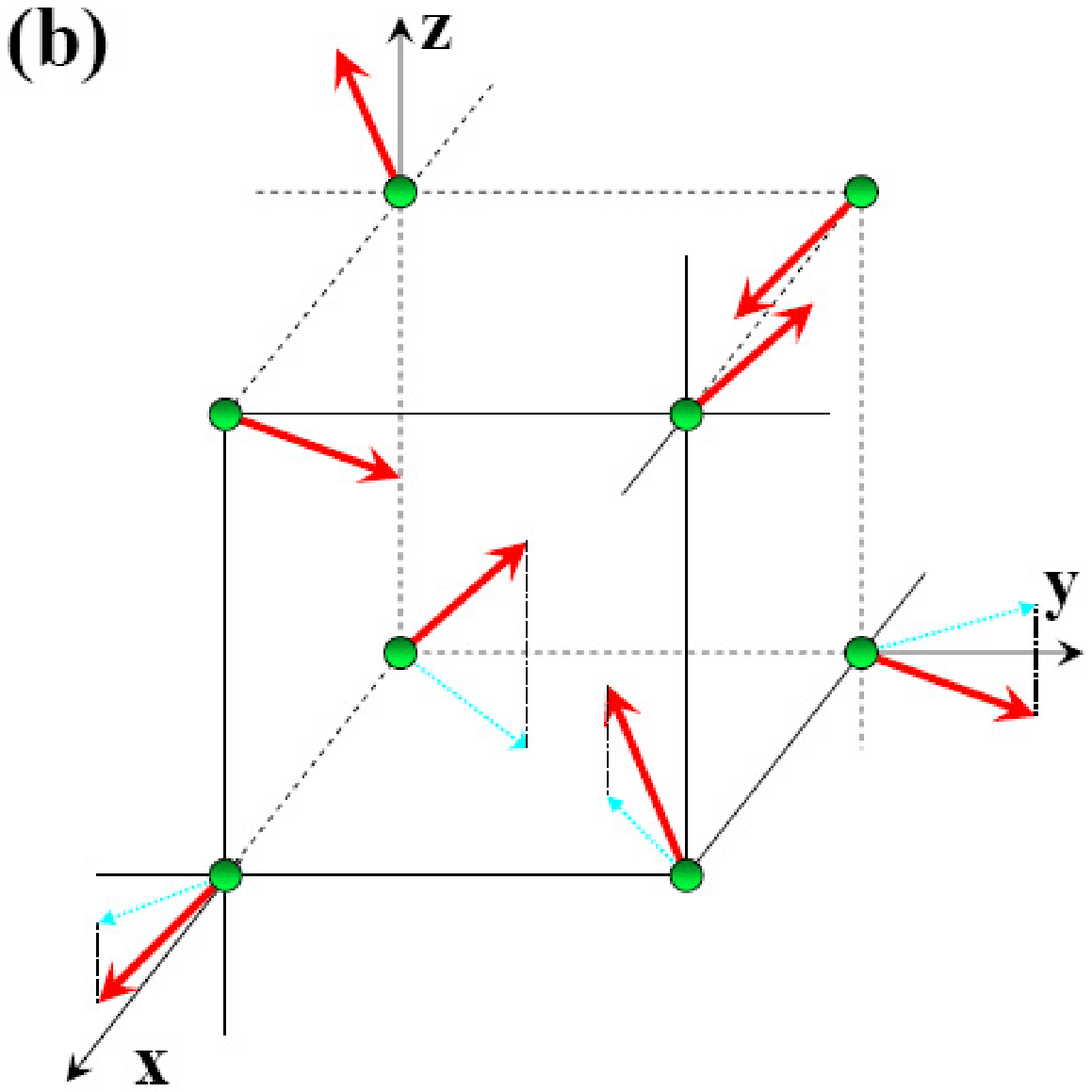}
\caption{(a) The SO(3) manifold of the 3D $p$-orbital BEC
order parameter in terms of Euler angles, where $\mathbf{l}$ is
the direction of the orbital angular momentum; (b) Sketch of the
non-collinear orderings of orbital angular momenta in real space
from Eq. (\ref{Eq:order}). }
\label{fig:Euler}
\end{figure}

We consider the orbital superfluid phase.
A remarkable feature of the band structure is that the energy minima of
$p_{\mu}~(\mu=x,y,z)$-orbitals are located at finite momenta $Q_\mu$ rather
than at zero momentum, which are 
$Q_x=(\frac\pi a,0,0); ~Q_y=(0,\frac \pi a,0);~Q_z=(0,0,\frac \pi a)$ 
for the three $p$-orbital subbands, respectively.
In the 3D cubic lattice, we will show that the onsite orbital angular momenta
are no longer collinear but exhibit orbital frustrations.
The single-particle states $\psi_{Q_\mu}=e^{iQ_\mu\cdot\mathbf{r}}$
($\mu=x,y,z$) are degenerate, thus any condensate wavefunction
of a linear superposition of these states,
\begin{equation}
|\vec{Q}\rangle=c_1|Q_x \rangle+c_2|Q_y \rangle+c_3|Q_z \rangle,
\label{Eq:tri}
\end{equation}
yields the same kinetic energy. The complex vector
$\vec{c}=(c_1,c_2,c_3)$ satisfies the normalization condition,
$|\vec{c}|^2=1$. Next we will consider the interaction effect to
lift the degeneracy and select the condensate wavefunctions.

\begin{figure}[htb]
\includegraphics[width=0.9\linewidth]{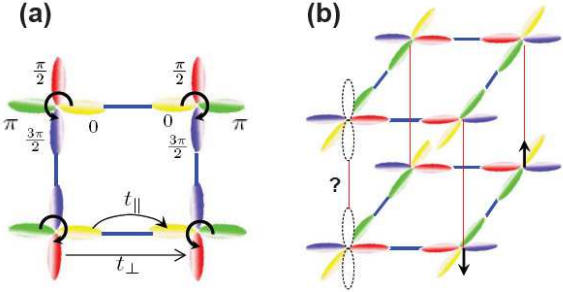}
\caption{(a) 2D orbital BEC in the square lattice without frustration.
(b) A typical configuration of 3D orbital BEC in a cubic lattice with
frustration. The thick blue bonds minimize both the transverse and
parallel hopping energy, while the thin red bonds only minimize the
transverse hopping energy.} \label{fig:frustrate}
\end{figure}

{\it The SO(3) degeneracy at the classical level}: At the classical
level (neglecting quantum fluctuations so that the boson operator
can be replaced by its average value), $H_{int}$ is minimized if the
coefficient vector $\vec{c}$ in Eq.(\ref{Eq:tri}) can be expressed
as $\vec{c}=\frac{1}{\sqrt 2} (\vec{m}+i\vec{n})$, where $\vec{m}$
and $\vec{n}$ are two mutual perpendicular unit vectors.
Transforming back into real space, for the lattice site with
integer-valued coordinates $\vec r=(r_x,r_y,r_z)$, its onsite
orbital configuration is $\frac{1}{\sqrt 2} (p_{\hat e_1}+ i p_{\hat
e_2})$ with the relation \bea
\hat e_1 &=& (P_x)^{r_x} (P_y)^{r_y} (P_z)^{r_z} \vec{m}; \nn \\
\hat e_2 &=& (P_x)^{r_x} (P_y)^{r_y} (P_z)^{r_z} \vec{n}, \eea where
$P_{x,y,z}$ are reflection operator with respect to $x,y,z$-axes,
respectively. $\hat e_{1,2}$ remain orthogonal to each other, and
the onsite orbital angular momentum $\vec L (\vec r)\pp \hat e_1
\times \hat e_2$, such that $\vec L^2$ is maximized to minimize
$H_{int}$. This denotes that at the classic level, the ground state
manifold is just the configuration space of the 3D orthogonal triad
$ \vec{m}, \vec{n}$ and $\vec{l}= \vec{m} \times \vec{n}$, which is
just the SO(3) group space and can be expressed in terms of Euler
angles ($\phi, \theta, \gamma$), as illustrated in
Fig.~\ref{fig:Euler}(a). Note that the multiplication of an overall
U(1) phase $e^{-i\varphi}$ is equivalent to the rotation of the
triad around $\vec{l}$ by the angle $\varphi$. Therefore, the U(1)
superfluid phase is absorbed into the SO(3) group configuration
space. For a given triad configuration $ \vec{m}_0$, $ \vec{n}_0$
and $ \vec{l}_0$, the corresponding real space distribution of the
OAM orientation $\hat L(\vec r)$ becomes \bea \hat
L_{\vec{r}}=[(-1)^{r_y+r_z}l_x,~(-1)^{r_x+r_z}l_y,~
(-1)^{r_x+r_y}l_z], \label{Eq:order} \eea which is non-collinear as
shown in Fig.~\ref{fig:Euler}(b).

Similarly to the case of frustrated magnets, this classic level degeneracy
is a consequence of orbital frustration, which means that it is impossible
to find an orbital configuration that simultaneously minimizes the energy
of all the bonds in the lattice.
To illustrate this point, let us recall the previously studied 2D
case for a comparison \cite{Liu2006}.
The staggered OAM configuration in Fig.~\ref{fig:frustrate} (a)
simultaneously minimizes both the parallel ($t_\|$) and transverse
($t_\perp$) hopping energies at all bonds of the square lattice,
and thus there is no frustration.
However, in the 3D cubic lattice, this is no longer the case.
For example, if we take a similar state $|Q_{xy}\rangle=\frac{1}
{\sqrt{2}}(|Q_x\rangle+i|Q_y\rangle)$, as shown in Fig.~\ref{fig:frustrate}(b),
the hopping energy of all bonds along $x$ and $y$ directions can be minimized,
but the $\sigma$-bond along the $z$-direction is broken.
Since the hopping Hamiltonian Eq. \ref{Eq:kin} does not preserve the
$SO(3)$ symmetry, thus this classic level degeneracy
should be lifted by quantum fluctuations.

{\it Order-from-disorder} In frustrated magnetism, the infinite degeneracy
is usually lifted by quantum or thermal fluctuations, which is
known as ``order-from-disorder'' mechanism \cite{Henley1989,Chubukov1992}.
Below we perform the same analysis to the 3D $p$-orbital BECs.
If we take quantum fluctuations around the mean-field values into account:
$p_\mu=\langle p_\mu\rangle+\delta p_\mu$ and calculate the
fluctuation-corrected ground state energy,  quantum
fluctuations lift the SO(3) classical degeneracy.
We consider two typical condensate configurations and compare their ground
state energies (the reason to choose these two states is due to
their high symmetry),
\bea
|Q_{diag}\rangle&=&\frac{1}{\sqrt{3}}(|Q_x\rangle+e^{i\frac
{2\pi}3}|Q_y\rangle+ e^{-i\frac {2\pi}3}|Q_z\rangle)\label{Eq:111},
\\
|Q_z\rangle&=&\frac{1}{\sqrt{2}}(|Q_x\rangle+i|Q_y\rangle).
\eea
In the state of  $|Q_{diag}\rangle$, OAMs are along the body-diagonal
directions, while for the state of $|Q_z\rangle$, OAMs are along the
$z$-direction.
These two configurations are degenerate at the classic level.

\begin{figure}[htb]
\includegraphics[width=0.8\linewidth]{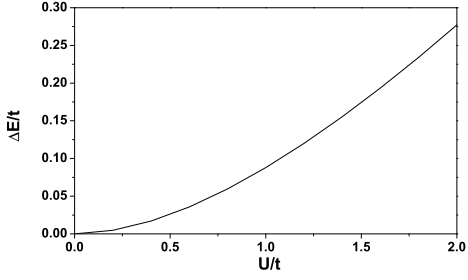}
\caption{The energy difference between the orbital BEC with OAM
towards $z$-direction  and body-diagonal direction $\Delta
E=E_z-E_{diag}$ with $t_\parallel=t$, $t_\perp=0.05t$ and the
filling factor $n$=2.} \label{fig:DeltaE}
\end{figure}

Here we perform the standard Bogoliubov analysis \cite{Oosten2001}
to calculate the zero-point motion energy of quasi-particles for these
two configurations. We use the state of $|Q_{diag}\rangle$ as an
example, and the calculation for $|Q_z\rangle$ is rather similar.
For each $p$-component the order parameter is
\bea
\langle p_x\rangle&=&(-1)^{r_x}\phi, \ \ \,  \ \ \, \langle
p_y\rangle=(-1)^{r_y}e^{\frac{2i\pi}3}\phi \nn \\
\langle p_z\rangle&=&
(-1)^{r_z}e^{-\frac{2i\pi}3}\phi. 
\label{meanfield}
\eea
To calculate the Bogoliubov spectra, we consider quantum
fluctuations around the mean-field values: $p_\mu=\langle
p_\mu\rangle+\delta p_\mu$. Expanding to the quadratic level, we
arrive at \bea i\hbar\frac{\partial \Psi(\mathbf{k}) }{\partial
t}=\mathcal{M}(\mathbf{k})\Psi(\mathbf{k}),\label{Eq:EOM} \eea where
$\Psi(\mathbf{k})$ is a 6-component vector that represents the
fluctuations:
$\Psi(\mathbf{k})=[\delta\psi(\mathbf{k}),\delta\psi^\dag(\mathbf{-k})]^T$,
$\delta\psi(\mathbf{k})=[\delta p_x(\mathbf{k}),\delta
p_y(\mathbf{k+Q_{xy}}),\delta p_z(\mathbf{k+Q_{xz}})]$,
$\mathbf{Q_{xy}}=(\pi,\pi,0)$, and $\mathbf{Q_{xz}}=(\pi,0,\pi)$.
$\mathcal{M}(\mathbf{k})$ is a $6\times6$  matrix as \bea
\nonumber\mathcal{M}(\mathbf{k})=\left [
\begin{array}{cc}
\mathcal{H}(\mathbf{k})& \Delta(\mathbf{k})\\
-\Delta^\dag(\mathbf{k})&-\mathcal{H}(\mathbf{-k})\\
\end{array}
 \right ],
\eea
in which both $\mathcal{H}$ and $\Delta(\mathbf{k})$ are $3\times3$ matrices:
\bea
\nonumber
\mathcal{H}(\mathbf{k})&=&\left [
\begin{array}{ccc}
\epsilon^x_{\mathbf{k}}+ 2w& -w&-w\\
-w&\epsilon^y_{\mathbf{k+Q_{xy}}}+2w &-w\\
-w&-w&\epsilon^z_{\mathbf{k+Q_{xz}}}+2w
\end{array}
\right ], \nn \\
\Delta(\mathbf{k})&=&\left [
\begin{array}{ccc}
w& e^{i\frac{2\pi}3}w& e^{-i\frac{2\pi}3}w\\
e^{-i\frac{2\pi}3}w& we^{-i\frac{2\pi}3}&w\\
w e^{-i\frac{2\pi}3}&w &w e^{i\frac{2\pi}3}
\end{array}
 \right ],
\eea where $w=\frac{2}{3} u \phi_{dg}^2$; $\epsilon^{\mu}_{\mathbf{
k}}=2\sum_\nu[t_\parallel\delta_{\mu\nu}-t_\perp(1-\delta_{\mu\nu})]\cos(k_\nu
a)$ is the single particle energy spectrum for the $\mu$ band boson.
The self-consistent equation to determine value of $\phi_{dg}$ is
\bea 
n=|\phi_{dg}|^2-\frac 12 +\frac 12
\sum_{\mathbf{k}}\frac{\bar{\varepsilon}(\mathbf{k})+2U|\phi_{dg}|^2}{\sqrt{\bar{\varepsilon}(\mathbf{k})(\bar{\varepsilon}(\mathbf{k})+4U|\phi_{dg}|^2)}}
\eea where
$\bar{\varepsilon}(\mathbf{k})=(\epsilon^x_{\mathbf{k}}
+\epsilon^y_{\mathbf{k+Q_{xy}}}+\epsilon^z_{\mathbf{k+Q_{xz}}})/3$,
and $n$ is the filling factor. 
The contribution from the zero point
motion energy to the ground state energy can be written as
\bea 
\nonumber E_{diag}^0=-3Un_c^2-Un_c-t+\frac 12
\sum_{\mathbf{k}}\sqrt{\bar{\varepsilon}(\mathbf{k})
(\bar{\varepsilon}(\mathbf{k})+4Un_c)}
\eea where $n_c=|\phi_{dg}|^2$, $t=t_\parallel +2t_\perp$.
Performing the same process, we obtain the correction for
$|Q_z\rangle$, and the difference $\Delta E=E_z-E_{diag}$ is plotted
in Fig.~\ref{fig:DeltaE}.

For fixed parameters $U, t_\parallel, t_\perp$ and boson density $n$, the
energy of $|Q_{diag}\rangle$ is always lower than that of $|Q_z\rangle$
i.e., $\Delta E=E_z-E_{diag}>0$, which means that orbital BECs in a cubic
lattice prefer to develop OAM moments along the body-diagonal directions.
Such a configuration has a high symmetry, and all the bonding strengths
are uniform in the lattice.
In comparison, all the $\sigma$-bonds along the $z$-direction are
broken in the state of $|Q_z\rangle$.
This ``order-from-disorder'' phenomenon is another feature that
distinguishes orbital BECs from spinor BECs.

\begin{figure}[htb]
\includegraphics[width=0.98\linewidth]{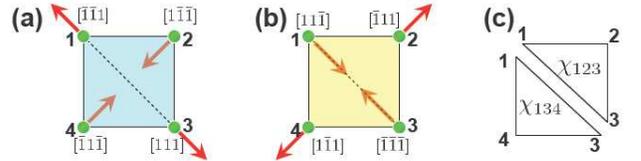}
\caption{Ordering patterns in a plaquette for (a) $|Q^1_{diag}\rangle$
and (b) $|Q^2_{diag}\rangle$ with opposite chirality; (c) Definition
of the chirality in a plaquette.} \label{fig:chiral}
\end{figure}

{\it Spontaneous chiral orbital order:} In condensed matter physics,
the presence of spin chirality plays important roles in frustrated
magnetism \cite{Grohol2005}, doped Mott-insulators \cite{Wen1989},
and the anomalous quantum Hall effect \cite{Taguchi2001}.
Here, we find that the orbital BECs in Eq.~(\ref{Eq:111}) spontaneously
develops a chiral orbital angular momentum ordering.
Eq. ~(\ref{Eq:111}) has a time-reversal partner as
\bea
|Q^\prime_{diag}\rangle=\frac{1}{\sqrt{3}}(|Q_x\rangle+e^{-i\frac
{2\pi}3}|Q_y\rangle+ e^{i\frac {2\pi}3}|Q_z\rangle).
\label{Eq:bbb}
\eea
The orbital angular momentum orderings of Eq. \ref{Eq:111} and
Eq. \ref{Eq:bbb} are plotted in Fig.~\ref{fig:chiral}.
To distinguish the chirality between $|Q_{diag}\rangle$ and
$|Q^\prime_{diag}\rangle$, we define a nonzero chirality
(similar to the case of chiral spin liquid \cite{Wen1989}):
\begin{equation}
\chi_{ijk}=\vec{l}_i\cdot (\vec{l}_j\times \vec{l}_k),
\end{equation}
where $ijk$ denotes three sites of the four corners of the plaquette
in a clockwise direction, as shown in Fig.~\ref{fig:chiral} (c). In
contrast to the chiral spin liquid \cite{Wen1989} (where $\langle
\vec{S}\rangle=0, \chi\neq 0$) and the spinor BEC (where $\langle
\vec{S}\rangle\neq 0, \chi=0$), the orbital BEC in our case
simultaneously exhibits non-vanishing orbital order and orbital
chirality order ($\langle \vec{l}\rangle\neq 0, \chi\neq 0$).

{\it Excitations:}  Now we discuss the collective modes and
elementary excitations of the orbital BECs in the state of
$|Q_{diag}\rangle$. From the EOM Eq.~(\ref{Eq:EOM}), we obtain the
collective modes by diagonalizing the matrix $\mathcal{M}$. In the
long-wavelength limit ($k\rightarrow 0$), we find that there are
three modes: (1) the gapless Goldstone mode corresponding to the
fluctuation of the superfluid phase of the orbital BECs with linear
dispersion, $\omega_1(\mathbf{k})\approx
\sqrt{2U|\phi|^2(2t_\perp+t_\parallel)}|\mathbf{k}|$, (2) the
orbital wave mode corresponding to the fluctuation of the OAM around
its ground state directions, $\omega_2(\mathbf{k})\approx
\frac{2t_\perp+t_\parallel}2k^2$, (3) the gapped mode, which
describes orbital excitations corresponding to the flipping of
orbital angular momenta, $\omega_3(\mathbf{k})\approx
\frac{2t_\perp+t_\parallel}2 k^2+2U|\phi|^2$.

{\it Ginzburg-Landau (GL) free energy:} To better understand the
various phases in orbital BECs, we first identify the order
parameters of the different phases. The orbital BECs are
characterized by both superfluidity with order parameters
$\varphi_\mu$ ($\mu=x,y,z$) and orbital order $\vec{L}$, which is a
3-component real vector $L_\mu$ (after the transformation defined in
Eq.~\ref{Eq:order}, the non-collinear order can be transformed into
a spatially uniform one). Notice that these two different orders are
not necessarily simultaneously present. To clarify this point, we
construct a GL free energy in terms of $\chi_\mu$ and $\vec{L}$:
\bea \nonumber \mathcal {F}&=&\sum_\mu
r|\varphi_\mu|^2+u|\varphi_\mu|^4+r^\prime|\vec{L}|^2+u^\prime(|\vec{L}|^2)^2
\nn \\
&+&\sum_{\mu\nu}v|\varphi_\mu|^2|\varphi_\nu|^2 -
g\varepsilon^{\lambda\mu\nu}i(\varphi_\mu^*\varphi_\nu-
\varphi_\nu^*\varphi_\mu)L_\lambda \label{Eq:GL}, \eea the last term
is the minimal coupling between the superfluid and orbital orders
and the parameters $g,~u',~u$ are positive in our case. For the the
orbital BEC in Eq.~\ref{Eq:111}, $\vec{L}$ is along the
body-diagonal direction and the free energy Eq.\ref{Eq:GL} can be
simplified to
\begin{equation}
\mathcal
{F}=r_0\phi^2+u_0\phi^4+r'_0l^2+u'_0l^4-g_0\phi^2l+\cdots.\label{Eq:free}
\end{equation}
The free energy $\mathcal {F}$ describes both thermodynamic phase
transitions which are driven by temperature and quantum phase
transitions where the parameters are a function of $U/t$. By
minimizing the free energy in Eq.~(\ref{Eq:free}), we find that: for
$r'_0<0$ and $r_0<\sqrt{-\frac{g_0^2r'_0}{2u'_0}}$, both $\phi$ and
$l$ are non-zero at the free energy minima, which corresponds to
orbital BECs with both superfluidity and orbital order; for $r'_0<0$
and $r_0>\sqrt{-\frac{g_0^2r'_0}{2u'_0}}$, $\phi=0$ while $l\neq 0$,
so in this case the single particle condensations are suppressed by
thermal or quantum fluctuations while the orbital order is
preserved; for $r'_0>0$ and $r_0>0$, both $\phi$ and $l$ are zero,
which means that both the superfluidity and orbital order have been
destroyed, corresponding to the  high temperature normal phases or
featureless Mott insulator at zero temperature.

{\it Experimental detection:} Next we discuss the experimental
detection of orbital ordering and unconventional BEC
characterized by the ordering parameters in Eq.(\ref{meanfield}).
The condensate wavefunction in Eq.(\ref{Eq:bbb}) is a superposition
of the single-particle states of three condensate momenta with equal
weights but different phases, as a consequence, the time-of-flight
(TOF) image will exhibits three peaks with the same height in the
points corresponding to $\mathbf{Q}_{x,y,z}$. 
However, TOF images only provide the single-particle density
distribution in momentum space, while the key information about the
relative phase  between different condensate components is lost
during TOF. 
To measure the phase difference $e^{\pm i\frac{2\pi}{3}}$ defined in 
Eq.(\ref{meanfield}), the phase sensitive detections proposed recently 
can be employed\cite{Cai2011a}.
Impulsive Raman operations can be applied to couple any two 
among the three components at different momenta.
The relative phase information can be read out from the 
interference pattern from the TOF imaging.

At last, we will briefly discuss the orbital BECs in higher  bands.
Recently, an unconventional BEC in the $f$-band of a bipartite
optical square lattice has been observed experimentally
\cite{Wirth2011a}. Surprisingly,  $d$-band BECs have also been
observed in a distinct field: the exciton-polariton condensate
\cite{Kim2011}. Orbital BECs in the $p$-band of 3D optical lattice
have three components ($p_x,~p_y,~p_z$), and the interactions favor
a ferro-orbital state with OAM $L=1$, which makes it similar to the
ferromagnetic phase of spinor BECs with $F=1$. Analogously, orbital
BECs in higher bands behave similarly to spinor BECs with higher
spin \cite{Barnett2006,Koashi2000,Ciobanu2000,Santos2006}.
Apparently for orbital BECs in higher bands, the geometry and
symmetry group of the order parameters is far more complex and may
give rise to richer physics.

In conclusion, we investigate the frustrated orbital ordering of
$p$-band unconventional BECs in the cubic lattice, and we find that the uniqueness
of the orbital degree of freedom gives rise to a lot of interesting
phenomena that absent in the spinor BECs and superfluid $^3$He, such
as orbital frustration and concomitant non-collinear orbital
orderings selected by the ``order-from-disorder" mechanism. The
chiral symmetry breaking and the elementary excitations have also
been discussed.

Z. C. and C. W. are supported by the NSF DMR-1105945 and the
AFOSR-YIP program. Y.W. gratefully acknowledges the support of Wuhan
University through the New Faculty Start-Up Grant.


\end{document}